\documentclass[aps,pre,twocolumn,superscriptaddress,showpacs,showkeys]{revtex4-1}
\usepackage{color}
\usepackage{amssymb}
\usepackage{amsmath}
\usepackage{newtxtext,newtxmath}
\usepackage{bm}
\usepackage{fixmath}
\usepackage{graphicx}
\usepackage{kotex}
\usepackage{xfrac}
\newcommand{\Corr}[1]{{\color{black}#1}}
\newcommand{\matr}[1]{\bm{{#1}}} 

\begin{document}
\setcounter{page}{1}
\title{Competition-induced increase of species abundance in mutualistic networks}
\author{Seong Eun \surname{Maeng}}
\author{Jae Woo \surname{Lee}}
\email{jaewlee@inha.ac.kr}
\author{Deok-Sun \surname{Lee}}
\email{deoksun.lee@inha.ac.kr}
\affiliation{Department of Physics, Inha University, Incheon 22212, Korea}

\begin{abstract}
Nutrients from a flowering plant are shared by its pollinators, giving rise to competition in the latter.  Such exploitative competition of pollinators can limit their abundance  and affect  the global organization of the mutualistic partnership in the plant-pollinator mutualistic community. Here we investigate the effects of the exploitative competition between pollinators on the structure and the species abundance of the mutualistic networks which evolve by changing mutualistic partnership towards higher abundance of species. Simulations show different emergent network characteristics between plants and animals; hub plants connected to many pollinators are very rare while a few super-hub pollinators appear with the exploitative competition included, in contrast to equally many hubs  of both types without the exploitative competition.  More interestingly, the abundance of plant species increases with increasing the exploitative competition strength. 
We analyze the inverse of the generalized interaction matrix in the weak-interaction limit to identify  the leading structural factors relevant to the species abundance, which  are shown to be instrumental in optimizing the network structure to increase the mutualistic benefit and lower the cost of exploitative competition.
\end{abstract}
\date{\today}
\maketitle 

\section{Introduction}
\label{sec:intro}

The organization of intra- and interspecific interaction strongly affects the stability of an ecological community and its response to perturbations~\cite{MAY:1972aa,Rohr1253497,Allesina:2015aa}. The ecological networks have been recently investigated to reveal their different structural characteristics depending on the nature of interactions and differentiating the dynamic behaviors of the corresponding communities~\cite{pascual05}. In case of flowering plants and animal pollinators that interact mutualistically, their network is characterized by high nestedness, meaning high likelihood of sharing mutualistic partners~\cite{bascompte03, almeidaneto08}. Such nested structure has been shown to enhance stability and biodiversity~\cite{bastolla09, thebault10,ELE:ELE12236}. On the other hand,  trophic networks develop modular structure to reduce competition in sharing a common prey~\cite{thebault10}.  

Structural characteristics serving for functional benefits such as high species abundance can be selected and enhanced during evolution. If certain organisms of a species form a mutualistic relationship with a new  partner and thereby achieve  a significantly high abundance and reproduction rate, then the species will be soon dominated by the descendants of these mutant organisms and their new partner will be among the list of mutualistic partners of the species. In accordance with the study revealing the advantage of the nested organization of mutualistic partnership for increasing the species abundance~\cite{bastolla09}, researchers have shown that the link rewiring conditioned towards increasing the species abundance can indeed generate highly nested structure in mutualistic networks~\cite{suweis2013}. 

Yet, in reality, various types of interactions may be present simultaneously  in a given community~\cite{Kefi:2016aa, Pilosof:2017aa}, which complicates the problem of what structural characteristics will emerge in the community. What interests us the most is that species in a mutualistic community can be subject to a competition similar to the one found in trophic networks; The benefit gained by an insect in pollinating a plant  can be reduced if there are many other pollinators occupying the same plant, as they should share its finite nutrient resource. This type of competition, which we call here exploitative competition, is different from the random interactions assumed between every pair of species in previous studies~\cite{PhysRevA.39.4333,0305-4470-22-17-011,PhysRevLett.93.178102} in that it exists only between the species sharing a common mutualistic partners. Therefore the organization of mutualistic relationship is coupled with that of exploitative competition.  The exploitative competition between pollinators sharing common plant partners has been incorporated in a growing network  model to explain the empirically observed asymmetric degree distributions between plants and animals~\cite{Maeng-Seong:2011aa,PhysRevLett.108.108701,PhysRevE.88.022804}. Compared with the introduction of new species, as considered in the growing network model,  a change in the mutualistic partnership can occur on a shorter time scale facilitating the enhancement or suppression of selected network characteristics to serve for increasing the species abundance.   

In this paper we investigate numerically and analytically the effects of the exploitative competition arising between the pollinators sharing the same plant species on the structure and the species abundance of the evolving plant-pollinator mutualistic network.  To this end, we use the co-evolution model~\cite{suweis2013} in which the abundance of each species is determined by the interspecific interaction encoded in the network structure and  the network evolves by  selective link rewiring towards abundance increase. The interaction matrix ruling the species abundance is here generalized to incorporate  the exploitative competition. 

We find that the iterative feedback between the species abundance and the network structure leads to  the optimal organization of mutualistic partnership that fully extracts the benefit of mutualism while lowers the cost of exploitative competition. Hub plant nodes are made rare, not to induce high competition between many pollinators sharing a plant. On the contrary, super-hub animals connected to all plants  appear, gaining large mutualistic benefit exceeding the exploitative competition cost at the expense of leaving many animals isolated. Interestingly, as the exploitative competition strength increases,  the abundance of animals fast decreases  but that of plants, counterintuitively, increases.  Given that the species abundance is determined by the inverse of the generalized interaction matrix in the co-evolution model, we expand it in the powers of interaction strength to identify the structural factors  relevant to the species abundance in the weak-interaction limit. The study of those leading  structural factors uncovers the combinatorial effects of the interspecific interactions of different nature, mutualism and competition,  on the evolution of ecological networks. 

The studied model is described in Sec.~\ref{sec:model} and the simulation results for the species abundance and the connectivity patterns emerging in the evolved network are presented in Sec.~\ref{sec:simul}. We derive and examine the analytic relation between the species abundance and the structural factors in Sec.~\ref{sec:analytic} and summarize our findings in Sec.~\ref{sec:summary}. 

\section{Model}
\label{sec:model}

 {\it Abundance dynamics with the intrinsic competition and mutualistic interaction--} 
We consider a model system consisting of  $N^{(P)}$ plant species and $N^{(A)}$ animal species. In the initial stage ($t=0$), the system has the interaction matrix $\matr{M}(0)$ mediating the quadratic intra- and interspecific interaction ruling the time-evolution of the abundance of each species $x_i$ as~\cite{bastolla09,suweis2013}
\begin{equation}
{dx_{i}\over d\tau}=\alpha_{i}\, x_{i} + x_i  \, \sum_{j=1}^S M_{ij}{(0)} \, x_j
\label{eq:dxdtau}
\end{equation}
 with $S=N^{(P)}+N^{(A)}$, $\tau$ the microscopic time, a much shorter scale than the macroscopic time or stage $t$, and $\alpha_i$   the intrinsic growth rate. The  $S\times S$ interaction matrix is  represented as 
\begin{equation}
\matr{M}(0)= -\matr{I} - 
\begin{pmatrix}
\matr{W}^{(P)} & \matr{0} \\
\matr{0} & \matr{W}^{(A)}
\end{pmatrix} + 
\begin{pmatrix}
\matr{0} & \matr{\Gamma}(0) \\
\matr{\Gamma^\intercal}(0) & \matr{0}
\end{pmatrix},
\label{eq:M0}
\end{equation}
 where the identity $I_{ij}=\delta_{ij}$ represents the unit rate of self-regulation. 
 The {\it intrinsic} competitions between the species of the same type, plant or animal, are encoded in the $N^{(P)}\times N^{(P)}$ matrix  $\matr{W}^{(P)}$ and the $N^{(A)}\times N^{(A)}$ matrix $\matr{W}^{(A)}$, the elements of which are random but fixed against time or stage;  $W^{(P)}_{pp'} (W^{(A)}_{aa'})$ is a positive random number  with probability $C$ or $0$ with probability $1-C$.  The mutualistic interaction between plant and animal species  is represented by the $N^{(P)}\times N^{(A)}$ matrix $\matr{\Gamma}(0)$, and its  elements are positive random numbers with probability $C$ or $0$ with probability $1-C$.  Here $C$  controls the fraction of interacting pairs, both mutualism and intrinsic competition, and will be called the connectance. 
 
  If there were no interspecific interaction $\matr{W^{(P,A)}}=\matr{0}$ and $\matr{\Gamma}(0)=\matr{0}$, Eq.~(\ref{eq:dxdtau}) would be reduced to a single-species equation $dx_i/d\tau =  \alpha_i \, x_i - x_i^2$, with which the  abundance $x_i$ converges to the only stable fixed point $x_i = \alpha_i$. This fixed point can be shifted by introducing non-zero interspecific interaction. We remark that  if Eq.~(\ref{eq:dxdtau}) contains a cubic term $x_i^3$, then two stable fixed points may emerge and the corresponding system becomes stable conditionally~\cite{Gao:2016aa}.

The mutualistic interaction matrix and the whole interaction matrix as well  evolve as stage $t$ changes, denoted by $\matr{\Gamma}(t)$ and $\matr{M}(t)$.  The mutualistic interaction is assumed to be symmetric such that $M_{ap}(t) = M_{pa}(t)$.  A bipartite network of $N^{(P)}$ plant-type nodes and $N^{(A)}$ animal-type nodes will be called the mutualistic network in this work, which has the $N^{(P)}\times N^{(A)}$ adjacency matrix $\matr{A}(t)$ with $A_{pa}{(t)}=1$ if $\Gamma_{pa}(t)>0$ and $0$ otherwise.   \Corr{Note that 
the interaction and adjacency matrices depend only on $t$ while the species abundance  is a function of both $t$ and $\tau$, denoted by $x_i (\tau;t)$ with $\tau$ running from $0$ to infinity for each stage $t\geq 0$. We will use simply $x_i$ instead of $x_i (\tau;t)$ and use $x_i(t)$ to represent the stationary abundance in each stage $\lim_{\tau\to\infty} x_i(\tau;t)$.} 

 {\it Inclusion of exploitative competition for $t\geq 1$--}
As introduced in Sec.~\ref{sec:intro}, we are particularly interested in the possibility that the mutualistic benefit  of a pollinator is reduced by the competition with other animal species pollinating the same plants. \Corr{To  quantify such reduction of benefit and incorporate it into the interaction matrix, we consider the effective mutualistic benefit and use it to construct the interaction matrix  for stage $t\geq 1$.  

For $t=1$, the mutualistic interaction matrix and the adjacency matrix are identical to those at $t=0$, i.e.,  $\matr{\Gamma}(1)=\matr{\Gamma}(0)$ and $\matr{A}(1)=\matr{A}(0)$, respectively.   In Eq.~(\ref{eq:dxdtau}) for $t=0$, we consider $M_{ap}(0) x_p$  as the mutualistic benefit of an animal species $a$ in pollinating a plant species $p$ and $M_{aa'}(0) x_{a'}$ as its cost of intrinsic competition with another animal species $a'$. They have the same dimension as  the growth rate $\alpha_a$, effectively modifying the latter.  Given that they are proportional to the abundance of mutualistic partner and other animal species, we can naturally assume that the reduction of mutualistic benefit caused by other pollinators at the same plant species  is proportional to their abundances, leading us to represent the effective mutualistic benefit of $a$ in pollinating  $p$ as}

\begin{equation}
M_{ap}{(1)} \, x_p =  M_{ap}{(0)} \, x_p- \sum_{a'}  \Corr{A_{p a} (1)} A_{p a'}{(1)} \ell_{aa';p}\, x_{a'},
\label{eq:benefit_single}
\end{equation}
where \Corr{$M_{ap}(0) = \Gamma_{pa}(0)=\Gamma_{pa}(1)$ and} $\ell_{aa';p}$ is the strength of the  competition between two animal species $a$ and $a'$ 
\Corr{in pollinating $p$. Summing Eq.~(\ref{eq:benefit_single}) over $p$, we find the whole benefit of $a$ from mutualistic interaction  to be represented as
\begin{align}
\sum_p M_{ap} {(1)} \, x_p = \sum_p M_{ap} {(0)} \, x_p - \sum_{a'} U_{a a'}(1) \, x_{a'},
\label{eq:benefit_all}
\end{align}
where 
\begin{align}
U_{a a'}(1) =  \sum_p A_{p a} (1) A_{p a'} (1) \ell_{a a';p} = k_{aa'}(1) \langle \ell_{aa';p}\rangle_p
\label{eq:U1}
\end{align}  
is  the sum of $\ell_{a a';p}$ over $k_{a a'}(1)\equiv\sum_p A_{p a}(1) A_{p a'}(1)$ distinct plant species co-pollinated  by $a$ and $a'$,  representing the strength of the   exploitative competition of $a$ and $a'$. 
Note that $\langle \ell_{aa';p}\rangle_p = \sum_p A_{pa}(1) A_{pa'}(1) \ell_{aa';p} /\sum_p A_{pa}(1) A_{pa'}(1)$ is the average of $\ell_{aa';p}$ over $p$, the co-pollinated plant species. Using Eq.~(\ref{eq:benefit_all}) in the time-evolution equation of the species abundance for $t=1$ like Eq.~(\ref{eq:dxdtau}), one finds that the whole interaction matrix is represented as
\begin{align}
\matr{M}(1) &= \matr{M}(0) - 
\begin{pmatrix}
\matr{0}& \matr{0} \\
\matr{0} & \matr{U}(1)
\end{pmatrix} \nonumber\\
&=
-\matr{I} +
\begin{pmatrix}
-\matr{W}^{(P)} &  \matr{\Gamma}(1) \\
\matr{\Gamma^\intercal}(1)  & -\matr{W}^{(A)} - \matr{U}(1)
\end{pmatrix}.
\label{eq:M1}
\end{align}


The mutualistic interaction $\matr{\Gamma}(t)$ and the adjacency matrix $\matr{A}(t)$ change with $t$ as will be detailed  below, and we constitute  the interaction matrix in the same form as in Eq.~(\ref{eq:M1}) not only for $t=1$ but for all $t\geq 1$:}
\begin{align}
\matr{M}(t) = 
-\matr{I} +
\begin{pmatrix}
-\matr{W}^{(P)} &  \matr{\Gamma}(t) \\
\matr{\Gamma^\intercal}(t)  & -\matr{W}^{(A)} - \matr{U}(t)
\end{pmatrix}.
\label{eq:Mt}
\end{align}
\Corr{Here, like Eq.~(\ref{eq:U1}),} the  exploitative competition matrix $\matr{U}{(t)}$  has elements
\begin{equation}
U_{a a'}{(t)}  = \rho\, u_{aa'} k_{aa'}{(t)},
\label{eq:U}
\end{equation}
where $k_{aa'}{(t)} = \sum_{p} A_{pa}{(t)}A_{pa'}{(t)}$, \Corr{ called here the overlap of $a$ and $a'$, and $\rho$ and  $u_{aa'}$ are  constants satisfying 
\begin{equation}
 \rho \, u_{a,a'} = \langle \ell_{a a';p}\rangle_p
 \end{equation}
under the assumption that the average $\langle \ell_{aa';p}\rangle_p$  is independent of time. The latter assumption is reasonable, considering that $k_{aa'}(t)$ will change more significantly with $t$ than $\langle \ell_{aa';p}\rangle_p$, which is an averaged quantity usually following a narrow distribution as stated in  the central limit theorem.}
$u_{aa'}$ is a random positive constant and $\rho$ is a control parameter. 
Note that $\matr{U}(t)$ depends on the mutualistic interaction $\matr{\Gamma}(t)$ or the adjacency matrix $\matr{A}(t)$ by Eq.~(\ref{eq:U}). 

{\it Co-evolution of the interaction matrix and the species abundance for $t\geq 2$ --} 
From Eq.~(\ref{eq:dxdtau}) with the interaction matrix $M_{ij}(t)$ used in place of $M_{ij}(0)$,  one can expect the  abundance \Corr{$x_i(\tau;t)$} of species $i$ to converge in the limit $\tau\to\infty$ to \Corr{$x_i(t)$} in stage $t$ given by
 \begin{equation}
\Corr{x}_i(t) = - \sum_{j=1}^N M^{-1}_{ij} (t) \, \alpha_{j}. 
\label{eq:stationary}
\end{equation}
This is expected to be stable against  slight perturbation \Corr{$\delta \matr{x} = \{ x_i(\tau;t) - x_i(t)\}$} as long as the magnitude of off-diagonal elements of $\matr{M}$ are sufficiently small~\cite{Allesina:2012aa,suweis2013,Allesina:2015aa}. Since a complicated $\tau$-dependent behavior of the species abundance in case of Eq.~(\ref{eq:stationary}) being unstable or in general cases~\cite{PhysRevE.95.042414} is not the main concern of the present study, we restrict out study to the case of weak interaction, i.e., the characteristic strength  of \Corr{intrinsic and exploitative} competition and mutualism  set to be small. Then, according to Eq.~(\ref{eq:stationary}), a change in the interaction matrix $\matr{M}$ may induce the increase or decrease of the species abundance. Given $\matr{M}(1)$ as in Eq.~(\ref{eq:Mt}) with $t=1$, the interaction matrix $\matr{M}(t)$ for $t\geq 2$ is generated \Corr{recursively} as follows.   At each stage $t$, we randomly choose a species $i$, plant or animal type, and change one of its partner $j$ by a new one $j'$ to form a trial mutualistic interaction matrix $\matr{\Gamma'}$ that is equal to $\matr{\Gamma}(t)$ except for the elements  $\Gamma'_{ij'}=\Gamma_{ij}(t)$ and $\Gamma'_{ij}=\Gamma_{ij'}(t)=0$ (or $\Gamma'_{j'i}=\Gamma_{ji}(t)$ and $\Gamma'_{ji}=\Gamma_{j'i}(t)=0$). Computing the trial exploitative competition matrix $\matr{U'}$ by using Eq.~(\ref{eq:U}) with $\matr{\Gamma'}$ and then the whole interaction matrix $\matr{M'}$ by Eq.~(\ref{eq:Mt}), we obtain the new abundance $x'_i$ of the selected species $i$ by Eq.~(\ref{eq:stationary}) with $\matr{M'}$. If $x'_i$ from $\matr{M}'$ is  not smaller than the original abundance $x_i(t)$ from $\matr{M}(t)$, then the considered link rewiring is accepted and we take $\matr{M'}$ as the interaction matrix at the next stage $t+1$, i.e., $\matr{M}(t+1)=\matr{M'}$. Otherwise, the link rewiring is disallowed and the  interaction matrix remains unchanged; $\matr{M}(t+1)=\matr{M}(t)$. To summarize, 
\begin{equation}
\matr{M}(t+1) = \left\{\begin{array}{ll}
\matr{M}' & {\rm if} \ x'_i \geq x_i(t), \\
\matr{M}(t) & {\rm otherwise}.
\end{array}
\right.
\label{eq:Mtp1}
\end{equation}
Such conditional rewiring is repeated every stage to generate a  series of  the interaction matrix $\{\matr{M}(t)|1\leq t\leq T\}$  and that of the  species abundance  $\{x_i(t)|1\leq i\leq S, 1\leq t\leq T\}$ with $T$ the total simulation stage.

The absolute value of a random number obeying the Gaussian distribution with zero mean and standard deviation $\sigma$ is assigned to $u_{aa'}$ for each pair of $a\neq a'$ in Eq.~(\ref{eq:U})  and to each non-zero positive element of $\matr{W^{(P,A)}}$ and $\matr{\Gamma}(1)$ in Eq.~(\ref{eq:Mt}).   Note that the elements of $\matr{\Gamma}(1)$ are continually exchanged with one another to form $\matr{\Gamma}(t)$ for $t\geq 2$. The parameter $\sigma$ characterizes the strength of  the intrinsic competition and  mutualism, and  $\rho \, \sigma$ is the strength of the exploitative competition. We choose a small value of $\sigma$ such that Eq.~(\ref{eq:stationary}) is a stable fixed point for Eq.~(\ref{eq:dxdtau}). The control parameter $\rho$ represents the characteristic relative strength of the exploitative competition compared to the mutualism or the intrinsic competition.  

While it is possible that the intrinsic growth rates $\alpha_i$'s are altered by environmental effects or various types of perturbations,  suggesting the volume of the 
$\alpha$ domain sustaining all species' survival as a stability measure~\cite{Rohr1253497}, we assume here that the mutations altering the intrinsic growth rate of a species occur on a much longer time scale than those leading to the change of its mutualistic partners. Also we assume that $\alpha_i$'s are not much different from species to species in a given mutualistic community.  
Given these considerations, we set  the growth rate to $\alpha_i = -\sum_j M_{ij}(0)$, which is close to $1$ as the off-diagonal terms of $\matr{M}$ are small, such that the abundance $x_i(0)$ at the initial stage ($t=0$) is $1$ for all $i$. This constitutes the initial condition of our model mutualistic network characterized by random interactions and an identical abundance for all species,  helping us to see clearly the co-evolution of network structure and species abundance. 


\section{Effects of exploitative competition: Simulation results}
\label{sec:simul}

\begin{figure}
\includegraphics[width=\columnwidth]{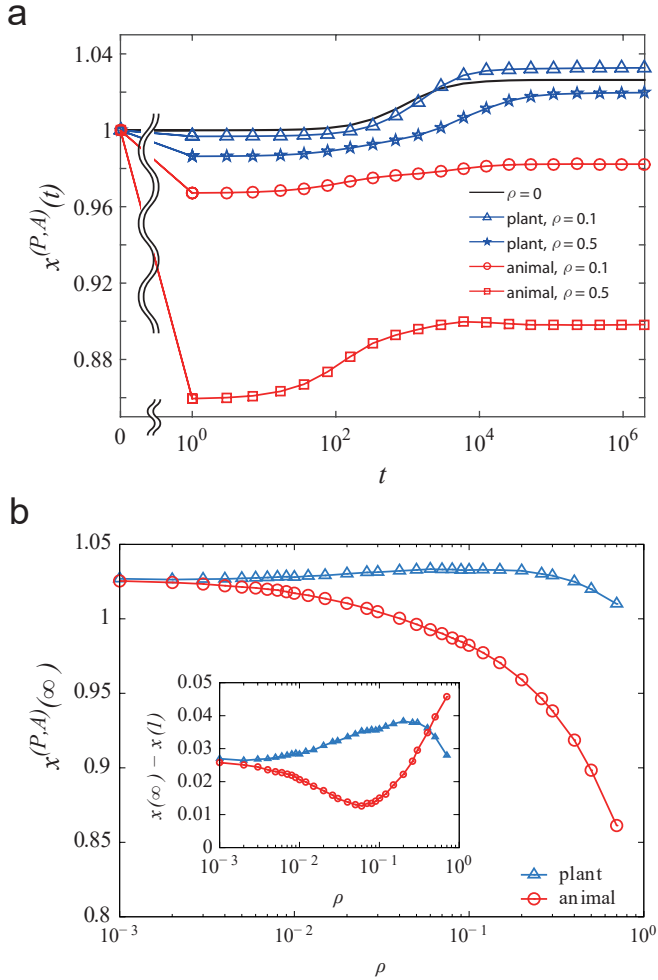}
\caption{The average abundance of a plant and an animal species. (a) The average abundance as a function of stage $t$ for different relative strength of exploitative competition $\rho$. 
(b) The stationary-stage abundance $x(\infty)$, evaluated by the ensemble average of the abundance in the final stage ($t=T$), as a function of $\rho$. Inset: The increase of the species abundance  $x(\infty)-x(1)$ during evolution for plants and animals plotted versus $\rho$.
}
\label{fig:Xpa}
\end{figure}

In this section  we present and analyze our simulation results for the evolution of the model mutualistic network for the  period $0\leq t\leq T=5\times 10^6$ with the number of species of each type $N^{(A)}=N^{(P)}=N=25$, the interaction strength $\sigma = 0.025$, the connectance $C=0.175$  following the empirical relation $C\simeq 4/S^{0.8}$ ~\cite{suweis2013}, and the relative strength of the exploitative competition $\rho$ varied between $0$ and $0.7$. Simulations with different numbers of species $N^{(A)}=N^{(P)} = 15, 40,50$, different numbers of species between  animal and plant, e.g., $\{N^{(A)},N^{(P)}\}=\{16,34\}$ or a different value of $\sigma=0.0125$ have been performed, but do not change qualitatively the results and discussions presented in this section. 

{\it Species abundance --} The average  abundance of a plant and an animal species increases as stage $t$ increases  after a drop at $t=1$  due to the inclusion of the exploitative competition for $t\geq 1$ [Fig.~\ref{fig:Xpa} (a)]. The abundance  in the stationary stage $t\to\infty$ shows big difference between plants and animals.  Animal's stationary-stage abundance $x^{(A)}(\infty)$ rapidly decreases with the exploitative competition strength $\rho$, but the abundance of  plant  $x^{(P)}(\infty)$ increases with increasing $\rho$ over a  range $\rho\lesssim  0.1$ [Fig.~\ref{fig:Xpa} (b)]. This anomalous increase of the plant abundance is interesting and must be related to the optimal structure of the mutualistic network obtained by evolution under different $\rho$'s. The increase of the species abundance $x(\infty)-x(1)$ during evolution is much larger in plants than animals for $\rho$ small as shown in the inset of Fig.~\ref{fig:Xpa} (b).

\begin{figure}
\includegraphics[width=\columnwidth]{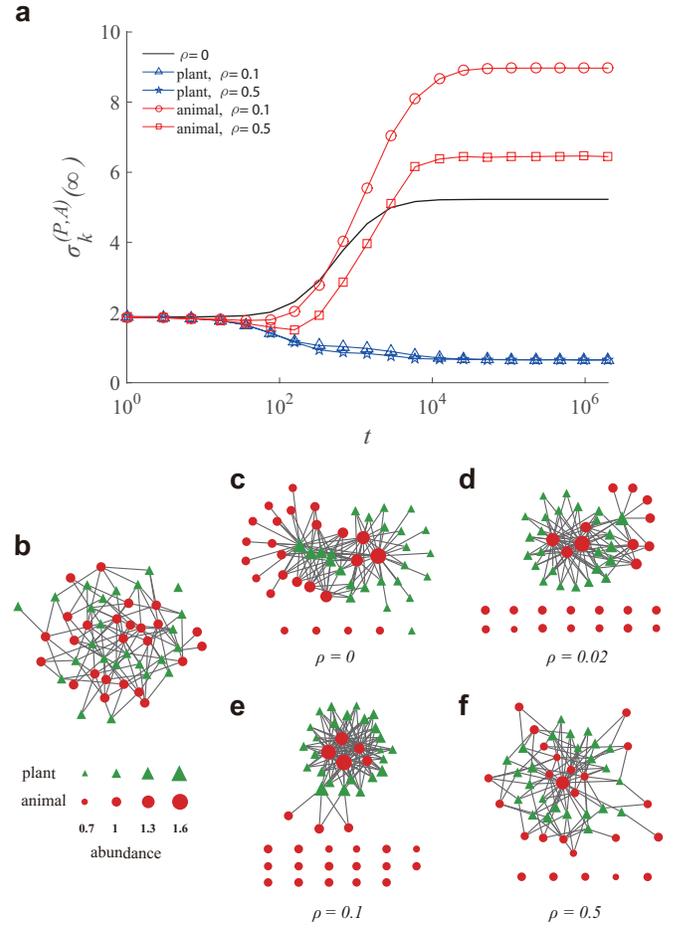}
\caption{Structural evolution of the mutualistic networks under exploitative competition. (a) Time-evolution of the standard deviation of node degree for different $\rho$'s. (b) An example of the mutualistic network in the initial stage $t=0$. It is a random bipartite network. Plant- and animal-type nodes are shown in green triangle and red circle, respectively.  (c-f) An evolved network in the final stage for (c) $\rho=0$, (d) $\rho=0.02$, (e)  $0.1$, and (f) $0.5$. Node size varies with the abundance. }
\label{fig:config}
\end{figure}

\begin{figure}
\includegraphics[width=0.95\columnwidth]{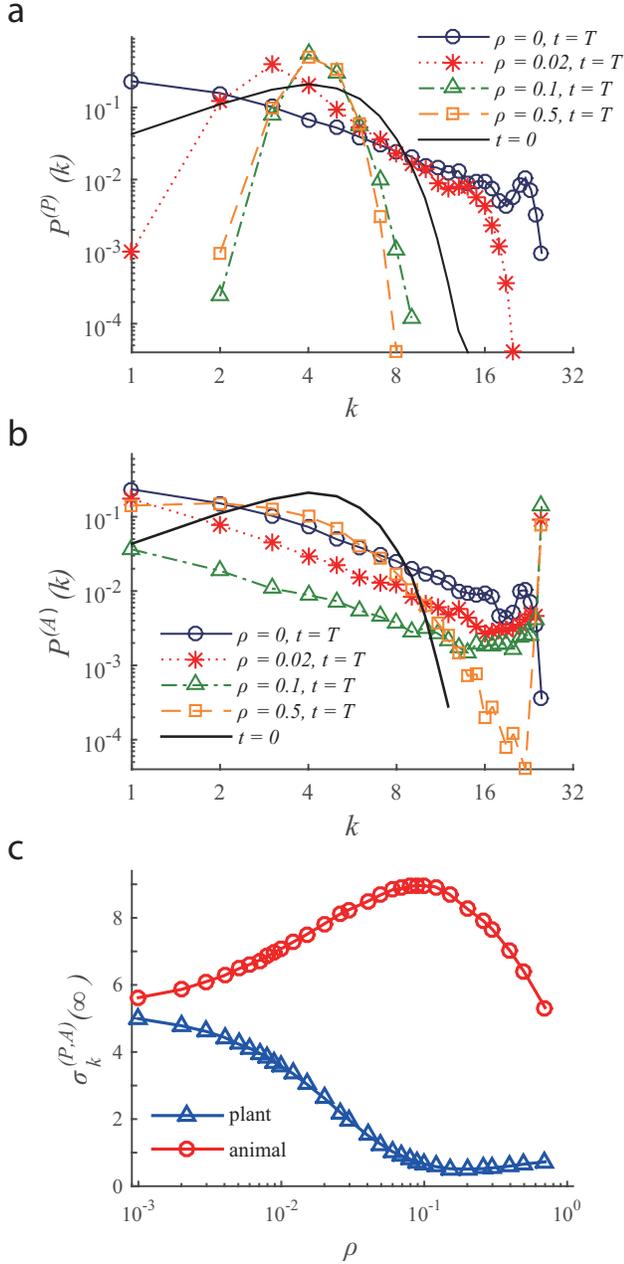}
\caption{Degree distribution of the evolving mutualistic networks. 
(a) The degree distribution $P^{(P)}(k)$ of plant nodes in the initial stage ($t=0$) and the final stage ($t=T$) for selected values of the exploitative competition strength $\rho$.
(b) The degree distributions $P^{(A)}(k)$ of animal nodes.   
(c) The standard deviations of the degree of plant and animal nodes in the final stage as functions of $\rho$.}
\label{fig:DD}
\end{figure}

{\it Network structure --} The standard deviation of the node degree varies  with time as shown in Fig.~\ref{fig:config} (a), implying a significant change in the network structure during evolution. The initial mutualistic network is a random bipartite network  [Fig.~\ref{fig:config} (b)].  In the stationary stage of evolution, the  mutualistic network comes to exhibit quite different characteristics from the initial network depending on $\rho$  as shown in Fig.~\ref{fig:config} (c-f). Without exploitative competition ($\rho=0$), a number of hub nodes of both plant and animal type exist, the abundance of which are high. The plant and animal hubs are  connected to each other and isolated nodes are very rare.   
Under the exploitative competition between animals ($\rho>0$), the  evolved network exhibits asymmetry between plants and animals; plant nodes are homogeneous whereas animal nodes are strongly heterogeneous in their connectivity pattern.  Hub plants or isolated plants are very rare. Plant nodes have similar degrees and abundance.  In contrast, a large number of animal nodes become isolated for $\rho>0$. Also there appear a few super-hub animal nodes that are connected to  all plant nodes.  Such heterogeneity of animal nodes is weakened when the exploitative competition is sufficiently strong ($\rho\gtrsim 0.5$) as shown in Fig.~\ref{fig:config} (f). 

\begin{figure}
\includegraphics[width=\columnwidth]{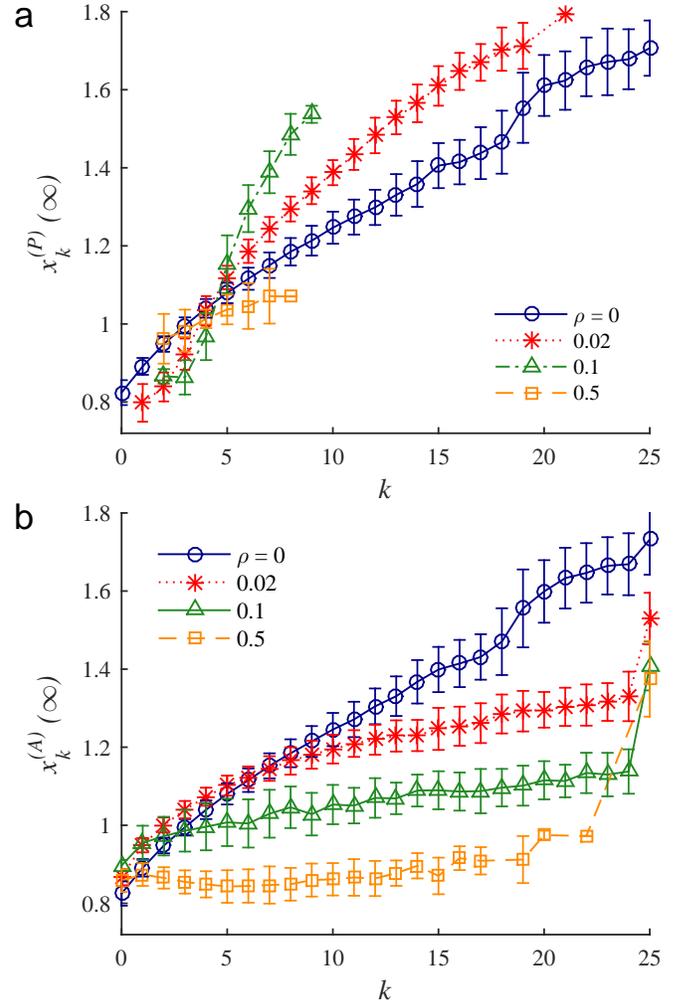}
\caption{The abundance of a species in the stationary stage as a function of its degree for selected values of $\rho$. (a) The average abundance $x^{(P)}_k(\infty)$ of a plant node having degree $k$. (b) The abundance $x^{(A)}_k(\infty)$ versus degree $k$. }
\label{fig:KXcorr}.
\end{figure}

{\it Degree distribution --} The structure of the mutualistic networks varying with $\rho$ can be quantitatively described in terms of the degree distribution. We find that the  degree  distribution in Poisson form at $t=0$ changes  to a  power-law  $P(k)\sim k^{-\gamma}$ with $\gamma\simeq 1.6$ in the stationary stage both for plants and animals with $\rho=0$ as shown in Fig.~\ref{fig:DD} (a) and (b).  With $\rho>0$, the degree distribution of plants is no more in  a power-law form; it gets narrower with increasing $\rho$ as shown in Fig.~\ref{fig:DD} (a) such that it is narrower even than the initial Poisson degree distribution for $\rho$ sufficiently large. The standard deviation of plants' degree  decreases with $\rho$ [Fig.~\ref{fig:DD} (c)], indicating that plants  have increasingly similar numbers of partners  with increasing $\rho$. On the other hand,   the animals' degree distribution  develops a peak  at the maximum possible degree $k=N^{(P)}$ [Fig.~\ref{fig:DD} (b)], indicating the appearance of the animal species pollinating all plant species. The appearance of such super-hub pollinators is made possible at the expense of leaving many animal nodes isolated. The presence of the animal nodes of such extreme connectivity, all or nothing,  leads to the large values of the standard deviation of the animals' degree in the range $\rho\lesssim 0.1$ [Fig.~\ref{fig:DD} (c)].  For $\rho\gtrsim 0.1$, even the degree distribution of animals loses its power-law form and becomes narrower with increasing $\rho$  [Fig.~\ref{fig:DD} (b) and (c)].

{\it Evolution strategy --} 
The large variation of the network structure with $\rho$ shown in Figs.~\ref{fig:config} (c-f) and \ref{fig:DD} reflects that the structural evolution strategy  for maximizing species abundance strongly depends on the exploitative competition. Furthermore,  the anomalous behavior of the  plants' abundance [Fig.~\ref{fig:Xpa} (b)] and the non-monotonic behavior of the standard deviation of the animals' degree [Fig.~\ref{fig:DD} (c)] suggest the nontrivial combinatorial effects of the mutualistic interaction and the exploitative competition on the optimal network structure. The abundance and the degree of individual species are in general positively correlated as shown in Fig.~\ref{fig:KXcorr}. Hub nodes benefit greatly from many partners  and thus their abundance can be large. Given that the preferential link attachment to nodes with many links can generate hubs and power-law degree distributions in growing networks~\cite{barabasi99,PhysRevLett.108.108701,PhysRevE.88.022804},  similar preferential link rewiring is expected to play a role during evolution to generate hubs and broad, power-law-like, degree distributions  for  $\rho=0$. Such heterogeneous connectivity pattern common to plant and animals is no more optimal if the exploitative competition is present. 
For $\rho>0$, if many distinct animal  species pollinated a common plant species, it would induce high exploitative competition between those pollinators and the abundance would be greatly reduced.  Therefore the pollinators are redistributed uniformly over as many plant species as possible, which is the origin of the narrow degree distribution of plants and the disappearance of isolated plant nodes. 
The number of  plant partners of an individual animal species is governed by  the benefit of mutualistic interaction  and the cost of exploitative competition. The super-hubs have their mutualistic  benefit exceeding well  the competition cost, due to the most partners possible. Their partners may have been taken from  small-degree animals seeing little possibility to have such high abundance as super hubs with their just few partners. For sufficiently large $\rho$, the cost of exploitative competition is so large that many pollinators come to have small degrees. The abundance of plant species stops to increase with $\rho$ but  decreases for such large $\rho$. 

These arguments help us understand how the exploitative competition induces  heterogeneity for animals  and homogeneity for plants  in the evolving mutualistic network.  However, they are still qualitative and it still remains open how much  a variation of structural characteristics changes the species abundance and what structural features are relevant to the abundance. The mechanism enabling  the plant abundance to increase with $\rho$ is in question. Quantitative answers to these questions can greatly advance our understanding of the interplay between the mutualistic benefit and the exploitative competition in the evolution of mutualistic community, which is the subject of the next section. 

\section{Structural factors relevant to species abundance: analytic approach}
\label{sec:analytic}

The mechanism optimizing the network structure for maximizing the species abundance  can be quantitatively understood by analyzing the relation between the species abundance and the interaction matrix in  Eq.~(\ref{eq:stationary}). The analysis can rely on the linearization or the expansion of the quantities of interest in powers of the characteristic strength $\sigma$ if  $\sigma$  is small. We first study  the accepted link rewiring to illustrate how the model network evolves with time and then obtain the stationary-stage abundance in terms of network properties.  

\subsection{Change of species abundance by  rewiring a link}

When a randomly selected species $i$ attempts to replace one of its partners by another,  this rewiring will be accepted if the abundance of $i$  is not decreased by this replacement, as in Eq.~(\ref{eq:Mtp1}).  In the weak interaction limit ($\sigma\to 0$), the resulting variation of the abundance $\delta x_i$ is related linearly to the change of the interaction matrix $\delta \matr{M}$ caused by the considered rewiring  as~\cite{suweis2013}
\begin{equation}
\delta x_i  =  \sum_j \delta M_{ij} \, x_j.
\label{eq:deltax}
\end{equation}

When a plant node $p$  replaces one of its partners $a_1$ by a new partner $a_2$,  the mutualistic interaction for the new pair  $(p, a_2)$ becomes $\Gamma_{p a_1}$  and  that for the old pair $(p,a_1)$ is changed to $0$. Using  $\delta M_{p a_1}=  -\Gamma_{p a_1}$ and $\delta M_{p a_2}= \Gamma_{p a_1}$,  one finds the variation of the abundance of $p$ given by 
\begin{equation}
\delta x_p =  (x_{a_2} - x_{a_1}) \Gamma_{p a_1}.
\label{eq:dxp}
\end{equation}
This rewiring will be accepted if the new partner $a_2$ is not less abundant than $a_1$.  Assuming a linear relation between the abundance and the degree of an animal species,  $x_a^{(A)} \simeq c^{(A)} \, k_a + b^{(A)}$ with $c^{(A)}$ and $b^{(A)}$ constants as suggested in Fig.~\ref{fig:KXcorr}  and taking the constant-interaction approximation  $\Gamma_{pa}=\sigma \, A_{pa}$, we obtain
\begin{equation}
\delta x_p \simeq \sigma\, c^{(A)} (k_{a_2} - k_{a_1}).
\label{eq:dxp2}
\end{equation}
This result shows the direction of the structural evolution:  a plant can increase its abundance by  replacing its animal partner by the one with a larger degree. Accordingly hub animal nodes can be created along with the broadened degree distribution [Fig.~\ref{fig:DD} (b)]. Such preferential rewiring is underlying the increase of the standard deviation $\sigma_k^{(A)}$ of animals' with the evolution stage  [Fig.~\ref{fig:config} (a)].

\begin{figure}
\includegraphics[width=\columnwidth]{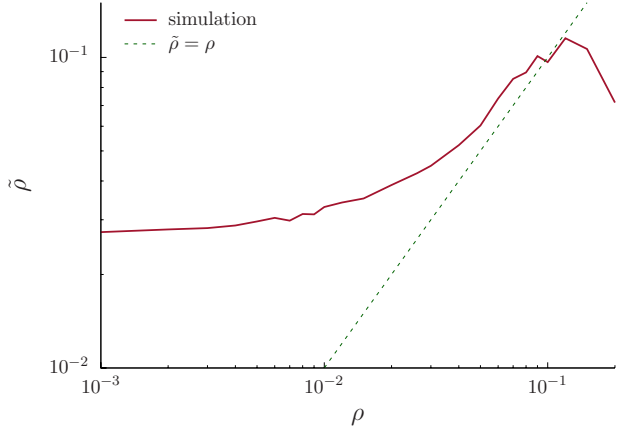}
\caption{Plot of $\tilde{\rho}$ in Eq.~(\ref{eq:tilderho}) as a function of $\rho$.  $\tilde{\rho}$ is computed by estimating the parameters $c^{(P)}, c^{(A)}, b^{(A)}$ and measuring $\sigma_k^{(A)}$ in simulation data. $\tilde{\rho}$ becomes smaller than $\rho$ for $\rho\gtrsim 0.1$. 
}
\label{fig:Tilderho}
\end{figure}

When an animal species $a$  replaces its plant partner $p_1$ by $p_2$,  not only the mutualistic benefit  but also the exploitative competition cost can be changed, resulting in the change of the abundance of $a$ given by 
\begin{equation}
\delta x_a = (x_{p_2} - x_{p_1})  \Gamma_{p_1 a}  - \sum_{a'}  (A_{a' p_2} - A_{a' p_1}) \Corr{\rho} \, u_{a a'} x_{a'}.
\label{eq:dxa}
\end{equation}
Introducing the approximation $x_p^{(P)} \simeq c^{(P)} \, k_p + b^{(P)}$ and  assuming  $\Gamma_{pa}=\sigma \, A_{pa}$ and $u_{aa'}=\sigma$, we have
\begin{equation}
\delta x_a \simeq  \sigma \left[ c^{(P)}  -\rho \langle x\rangle'\right]   (k_{p_2}-k_{p_1}),
\label{eq:dxa2}
\end{equation}
where  $\langle x\rangle' \equiv c^{(A)} \langle k\rangle' + b^{(A)}$ and $\langle k\rangle' \equiv {\sum_{p}\sum_a A_{pa} k_a \over \sum_{p }\sum_a A_{pa}} = {\sum_a k_a^2 \over \sum_a k_a}$ are the average abundance and the degree, respectively, of a pollinator connected to a randomly-selected plant and $\sum_{a} A_{p a} k_{a} \simeq k_{p} \langle k\rangle'$ is assumed for any $p$. Let us define $\tilde{\rho}$ as
\begin{equation}
\tilde{\rho} \equiv {c^{(P)} \over \langle x\rangle'}.
\label{eq:tilderho}
\end{equation}  
Eq.~(\ref{eq:dxa2}) indicates that animal nodes will tend to replace their partners by the ones having large (small) degrees, broadening (narrowing) the degree distribution of plants  if $\rho<\tilde{\rho}$ ($\rho>\tilde{\rho}$). The inequality between $\rho$ and $\tilde{\rho}$ should be related to which is dominant of  the mutualistic benefit and the exploitative competition cost. By using the estimated parameters and the second moments from simulation results, we can evaluate $\tilde{\rho}$ as a function of $\rho$, which is shown in Fig.~\ref{fig:Tilderho}.   It turns out that $\rho$ remains smaller than $\tilde{\rho}$ for $\rho\lesssim 0.1$, which explains why the degree distribution of plants in the stationary stage is broader for $\rho=0.02$ but narrower with $\rho=0.1$ or  $0.5$ than the initial Poisson distribution in Fig.~\ref{fig:DD} (a). The standard deviation $\sigma_k^{(P)}(\infty)$ of plants' degree decreases with $\rho$ passing  the initial-stage value $\sigma_k^{(P)}(0) \simeq 2.1$ around $\rho \simeq 0.03$ which is comparable to the threshold $0.1$ at which $\tilde{\rho}=\rho$.  

Although obtained by approximations, Eqs.~(\ref{eq:dxp2}) and (\ref{eq:dxa2}) explain well the evolution of the connectivity pattern of the mutualistic network; The degree distribution of animals is broadened  for all $\rho$ while that of plants is broadened only for $\rho$ small.  

\subsection{Structural factors mediating direct and indirect interactions}

\begin{figure}
\includegraphics[width=\columnwidth]{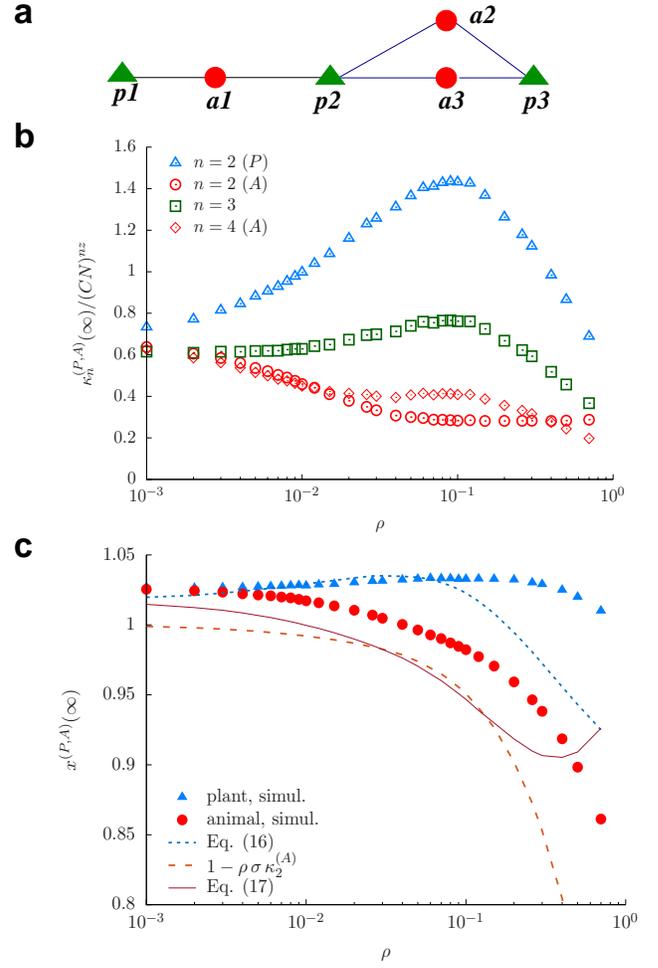}
\caption{Characteristic structural factors relevant to the species abundance. (a) In this example bipartite network of $3$ plant and animal nodes, two distinct  paths of length $2$ connect $p_2$ and $p_3$   and only one path of length $3$ connect $p_1$ and $a_3$. Using the adjacency matrix $\matr{A} = \left( \begin{array}{ccc}1 & 0 & 0\\ 1  &1 & 1 \\ 0 & 1 &1\end{array}\right)$ in Eq.~(\ref{eq:kappa}), one finds that $\kappa_2^{(P)}=4, \kappa_2^{(A)}=14/3, \kappa_3 = 28/3,  \kappa_4^{(A)}=22$. 
(b) Characteristic structural factors $\kappa_n$'s in the evolved networks.  They are rescaled by $(CN)^{nz}$ with $z=1.44$ selected just for the best data collapse at $\rho=0$. (c) Species abundance (lines) obtained by inserting the measured values of $\kappa_n$'s  into Eqs.~(\ref{eq:xp}) and (\ref{eq:xa}). Directly measured abundances (triangles, circles) are shown for comparison.  
}
\label{fig:theory}
\end{figure}

Here we obtain the inverse of the interaction matrix $\matr{M}^{-1}$ up to the second order of $\sigma$, which reveals all  direct and indirect interactions between two species connected by a link or a path of length two. The mechanism enabling the anomalous increase of the plant abundance under increasing exploitative competition will be illuminated by this analysis.

Let us take the constant-interaction approximation for the intrinsic competition,  $W_{ij}^{(P,A)} = C\sigma (1-\delta_{ij})$ for all $i$ and $j$ as well as for the mutualism and the exploitative competition, $\matr{\Gamma} =\sigma \matr{A}$, and $\matr{U}=\rho\, \sigma \matr{A^\intercal} \matr{A}$. Then  the sum of the self-regulation and the  intrinsic competition $\matr{M}_0 \equiv -\matr{I}-\left( \begin{array}{cc} \matr{W}^{(P)} & 0 \\ 0  &\matr{W}^{(A)} \end{array}\right)$ has its inverse exactly obtained as $\matr{M}_0^{-1} = - \left( \begin{array}{cc} \matr{G}_0 & 0 \\ 0  &\matr{G}_0 \end{array}\right)$ with~\cite{suweis2013}
\begin{equation}
(G_0)_{ij}= {\delta_{ij} \over 1-C\sigma}   - {C\sigma\over (1-C\sigma) (1+C\sigma(N-1))}.
\label{eq:G0}
\end{equation}
 The inverse of the whole interaction matrix $\matr{M}^{-1} = (\matr{M}_0+ \matr{V})^{-1}$ can be expanded in terms of  the network-dependent  interaction $\matr{V} \equiv \sigma \left( \begin{array}{cc} 0 & \matr{A} \\ \matr{A^\intercal}  &\rho  \matr{A^\intercal A} \end{array}\right)$ as $\matr{M}^{-1} = \matr{M}_0^{-1} - \matr{M}_0^{-1} \matr{V} \matr{M}_0^{-1} + \matr{M}_0^{-1}\matr{V} \matr{M}_0^{-1} \matr{V} \matr{M}_0^{-1} + O(\matr{V}^3)$. 
Up to the second order of $\sigma$, the inverse $\matr{M}^{-1}$ is represented as 
\begin{align}
- &\matr{M}^{-1} =
 \left( \begin{array}{cc} \matr{G}_0 & \matr{0} \\ \matr{0}  &\matr{G}_{0} \end{array}\right) + 
  \sigma \left( \begin{array}{cc} \matr{0} & \matr{G}_0 \matr{A} \matr{G}_0 \\  \matr{G}_0 \matr{A^\intercal} \matr{G}_0 &-\rho \, \matr{G}_0 \matr{A^\intercal A}\matr{G}_0\end{array}\right)\nonumber\\
 &+\sigma^2 \left( \begin{array}{cc} \matr{AA^\intercal} & -\rho \matr{A A^\intercal A} \\ -\rho \matr{A^\intercal A A^\intercal}  &\matr{A^\intercal A} + \rho^2 \matr{A^\intercal A A^\intercal A} \end{array}\right) + O(\sigma^3).
\label{eq:M-1}
\end{align}
Using Eqs.~(\ref{eq:G0}) and (\ref{eq:M-1}) into Eq.~(\ref{eq:stationary}) with  $\alpha_j = 1$,  we obtain the average abundance of plants 
\begin{align}
x^{(P)} \simeq   1 + \sigma^2 \left\{  \kappa_2^{(P)} - (C N)^2   -\rho \, \kappa_3\right\},
\label{eq:xp}
\end{align}
and  the average abundance of animals 
\begin{align}
x^{(A)} &=  1 - \rho\,  \sigma \, \kappa_2^{(A)} \nonumber\\
&+ \sigma^2\left\{
   (1+ 2\rho\, (CN))   \kappa_2^{(A)} -(C N)^2 
-\rho \,  \kappa_3 + \rho^2\, \kappa_4^{(A)}\right\},
\label{eq:xa}
\end{align}
where $N$ is the number of plant or animal species set equal for simplicity. Note that the benefit of the direct mutualistic interaction $\sigma \sum_{pa} (\matr{G}_0 \matr{A} \matr{G}_0)_{pa}\simeq CN\sigma$ in the first order of $\sigma$ is canceled out by the intrinsic competition included in  $\sum_p (G_0)_{pp}= {1\over C\sigma (N-1)}\simeq 1- CN \sigma$. On the other hand, the exploitative competition survives  in the $\sigma$ order in the animal's abundance.   In these expansions are $\kappa_n$'s with $n=2,3,4$, which are characteristic structural factors mediating the direct or indirect interspecific interaction and defined as
 \begin{align}
\kappa_2^{(P)} &= {1\over N} \sum_{p,p'} (\matr{A}\matr{A}^\intercal)_{pp'}, \ \kappa_2^{(A)} = {1\over N} \sum_{a,a'} (\matr{A}^\intercal \matr{A})_{aa'}, \nonumber\\
\kappa_3 &= {1\over N} \sum_{p,a} (\matr{A}\matr{A}^\intercal \matr{A})_{pa} ={1\over N} \sum_{a,p} (\matr{A^\intercal A A^\intercal})_{ap}, \nonumber\\
\kappa_4^{(A)}  &= {1\over N} \sum_{a,a'} (\matr{A^\intercal A A^\intercal A})_{aa'},
\label{eq:kappa}
\end{align}
with examples in Fig.~\ref{fig:theory} (a). The $\kappa_n^{(P,A)}$ indicates  the total number of length-$n$ paths connecting two nodes of given type. Notice that  $\kappa_2^{(A)}$ is the average overlap between animals $\kappa_2^{(A)} = N^{-1}\sum_{aa'} k_{aa'}$ with $k_{aa'}$  in Eq.~(\ref{eq:U}). Also $\kappa_2$ is directly related to the standard deviation of degree, i.e.,  $\kappa_2^{(P)} = \left(\sigma_k^{(A)}\right)^2 + \langle k\rangle^2$ and $\kappa_2^{(A)} = \left(\sigma_k^{(P)}\right)^2 + \langle k\rangle^2$, and similar to the nestedness~\cite{almeidaneto08}  except that the overlap is summed  only over the pairs of nodes having different degrees for the nestedness.

The  $\kappa_n$'s measured in simulation data are shown in Fig.~\ref{fig:theory} (b). In the range $0<\rho\lesssim  0.1$, $\kappa_2^{(P)}$ and $\kappa_3$ increase  while $\kappa_2^{(A)}$ and $\kappa_4^{(A)}$ decrease. For $\rho\gtrsim 0.1$, those  factors do not change or decrease with $\rho$. Inserting those measured values of  $\kappa_n$'s  into Eqs.~(\ref{eq:xp}) and (\ref{eq:xa}), we obtain the species abundance in good qualitative agreement with those of the directly measured abundance  for $\rho$ small, less than $0.1$, as shown in Fig.~\ref{fig:theory} (c). In particular, the increasing abundance of plants with $\rho$ is reproduced also by Eq.~(\ref{eq:xp}) despite significant deviation for $\rho\gtrsim 0.1$.

If there is no exploitative competition ($\rho=0$), the larger the  average overlap $\kappa_2$ is, the larger the abundance of plants and animals is according to Eqs.~(\ref{eq:xp}) and (\ref{eq:xa}).  The $\kappa_2^{(P)} (\kappa_2^{(A)})$ term appears in the second order of $\sigma$, as it represents an indirect interaction between plants (animals) arising from their respective mutualistic interactions with common pollinators (plants). This distance-2 mutualistic interaction obviously makes a positive contribution to the species abundance, underlying the well-known positive correlation between the nestedness and the species abundance~\cite{bastolla09,suweis2013}. 
If $\rho>0$, $\kappa_2^{(A)}$ is utilized also for the exploitative competition between animals which reduces directly the animal abundance in the first order of $\sigma$ in Eq.~(\ref{eq:xa}).  $\kappa_2^{(A)}$ should be therefore made small to reduce the decrease of the animal abundance.   Animals' stronger preference of plant partners with small degrees during evolution, as shown in Eq.~(\ref{eq:dxa2}), can bring smaller $\kappa_2^{(A)}$ and smaller $\sigma_k^{(P)}$ in the stationary stage for larger $\rho$. Nevertheless the abundance of animals cannot help but decrease with $\rho$  [Fig.~\ref{fig:theory} (c)]. 

The exploitative competition affects the plant abundance only indirectly;  The $\kappa_3$ term in Eq.~(\ref{eq:xp}) represents the indirect interaction between a plant $p$ and an animal species $a$,  having another animal species $a'$ interacting mutualistically with $p$ and competing exploitatively with $a$. The abundance of $a$ is thus detrimental to $p$ in this example.  As long as $\rho\lesssim 0.1$, such negative effect of the next-nearest neighbors on the plant abundance in the $\sigma^2$ order is not so strong as the positive effect of the distance-2 mutualistic interaction mediated by $\kappa_2^{(P)}$, which  enables  the anomalous increase of $x^{(P)}(\infty)$ with $\rho$.  While $\rho$ increases to $0.1$, $\kappa_2^{(P)}$ of the evolved networks increases significantly overcoming the increasing negative effect $-\rho \kappa_3$ given in Eq.~(\ref{eq:xp}). See Fig.~\ref{fig:theory} (b).  As the networks with many hub nodes tend to have high nestedness~\cite{Lee:2012aa},  plants' preferential partnering with animals of large degrees, indicated by  Eq.~(\ref{eq:dxp2}), is expected to be  the mechanism used to increase $\kappa_2^{(P)}$ and equivalently $\sigma_k^{(A)}$ during evolution. 

These results suggest that $\kappa_2^{(P)}$ and $\kappa_2^{(A)}$ are the key driver of the structural evolution, controlling the distance-2 indirect mutualistic interaction benefit and the direct exploitative competition cost. For $0<\rho\lesssim 0.1$, $\kappa_2^{(A)}$ is made small whereas $\kappa_2^{(P)}$ is made large during evolution, which is the evolution strategy to maximize the species abundance. Moreover, the increase of $\kappa_2^{(P)}$ with $\rho$ underlies the counterintuitive increase of the plant abundance. 

When $\rho$ is sufficiently large ($\rho\gtrsim 0.1$),  the increasing effect of the indirect interaction $-\rho \kappa_3$ on the plant's abundance is significant, which can be a reason for the decrease of $\kappa_3$ with  $\rho$. As $\kappa_n$'s are just different powers of the same adjacency matrix, the decrease of $\kappa_2^{(P)}$ with $\rho$  may be related to the decrease of $\kappa_3$. $\kappa_2^{(A)}$ remains almost the same while $\kappa_4^{(A)}$ decreases with $\rho$. It should be noted that $\kappa_3$ and $\kappa_4$  can contribute to the species abundance in the $\sigma^3$ and $\sigma^4$ order, beyond the scope of Eqs.~(\ref{eq:xp}) or (\ref{eq:xa}). Moreover, the abundance of plants for large $\rho$  remains much larger than the prediction by Eq.~(\ref{eq:xp}), implying the non-negligible roles of higher-order structural features.

\section{Summary and discussion}
\label{sec:summary}

In this work, we investigated how the plant-pollinator mutualistic partnership  is organized to increase the species abundance given the exploitative competition between the  pollinators  sharing the same plant  partners. The relevance of such exploitative competition to the structure of mutualistic networks has been implied previously but its effects on the  network evolution coupled with the species abundance have been first investigated here.  The most interesting finding is that the exploitative competition between pollinators results in an anomalous increase of the plant abundance while the animal abundance is decreased.  The evolutionary pressure is imposed in the direction of making the connectivity of plants homogeneous and that of animals heterogeneous.  Expanding the inverse of the interaction matrix in the weak-interaction limit uncovers the first few leading direct and indirect interactions affecting the species abundance, which allows us to find that the average overlaps between plants and between animals are instrumental for the structural optimization to achieve the maximum possible species abundance in the presence of both mutualism and the exploitative competition. 

Our study demonstrates that a structural factor can be enhanced or suppressed during evolution depending on the interplay between different types of interspecific interactions, emphasizing its importance in understanding  the structure and function of an ecological community. Given that the indirect mutualistic interaction and the exploitative competition drive the overlap or the nestedness  in the opposite directions,  it is  desirable to develop a method to infer the relative strength of the exploitative competition of real-world mutualistic networks from the empirical data. 
\Corr{Also, asymmetry in the overlap or nestedness between plants and pollinators can be measured empirically and its comparison across different  mutualistic communities may inform us of different  strengths of exploitative competition across communities. When the empirical data of the overlap and the species abundance of plants and pollinators are available, their relation can be checked against the theoretical prediction presented in this work.}

\begin{acknowledgments}
We thank Samir Suweis for helpful comments and Sung-Gook Choi for visualizing network evolution. This work was supported by the National Research Foundation of Korea (NRF) grants funded by the Korean Government (Grants No. 2017R1A2B4005214 (JWL) and 2016R1A2B4013204 (DSL)).
\end{acknowledgments}


%

\end{document}